\documentclass{raa}            % referee version: for submission

\usepackage{graphicx,times}             %for PS/EPS graphics inclusion, new
\usepackage{natbib}
\usepackage{amssymb,amsmath}
\bibpunct{(}{)}{;}{a}{}{,}

\usepackage[a4paper=true,pagebackref=true]{hyperref}
\hypersetup{colorlinks = true, linkcolor = green, anchorcolor = red, citecolor = blue, filecolor = red, pagecolor = red, urlcolor = red}

\begin{document}

   \title{The formation of blue cluster in local Universe
%\footnotetext{$*$ Supported by the National Natural Science Foundation of China.}
}
%   \subtitle{I. Place Your Subtitle Here}

   \volnopage{Vol.0 (20xx) No.0, 000--000}      %%preserved for Editor. DOn't remove!
   \setcounter{page}{1}          %%starting page, preserved for Editor. DOn't remove!

   \author{Qingxin Wen
      \inst{1,2}
   \and Yu Luo
      \inst{1,2}
   \and Xi Kang
      \inst{3,1}
   }

   \institute{Purple Mountain Observatory, Chinese Academy of Sciences, 10 Yuanhua Road, Nanjing 210033, China; {\it luoyu@pmo.ac.cn}\\
        \and
            School of Astronomy and Space Sciences, University of Science and Technology of China, Hefei 230026, China; \\
            \and
            Zhejiang University-Purple Mountain Observatory Joint Research Center for Astronomy, Zhejiang University, Hangzhou 310037, China; {\it kangxi@zju.edu.cn} \\
\vs\no
   {\small Received~~20xx month day; accepted~~20xx~~month day}}

\abstract{It is well known from the Butcher-Oemler effect that galaxies in dense environment are mostly red with little star formation and the fraction of blue galaxies in galaxy groups/clusters also declines rapidly with redshifts. A recent work by Hashimoto et al. reported a local 'blue cluster' with high fraction of blue galaxies ($\sim 0.57$), higher than the model predictions. They ascribed this blue cluster to the feeding of gas along a filamentary structure around the cluster.  In this work we use group catalog from the Sloan Digital Sky Survey Data Release 7 (SDSS DR7) and the state-of-art of semi-analytic model (SAM) to investigate the formation of blue clusters in local Universe.  In total, we find four blue clusters with halo mass  $\sim 10^{14}M_{\odot}$ at $0.02 < z < 0.082$, while only the one found by Hashimoto et al. is in a filamentary structure. The SAM predicts that blue clusters have later formation time and most blue satellite galaxies are recently accreted.  We conclude that the formation of blue clusters is mainly governed by newly accreted blue satellites, rather than the effect of large-scale environment.
}

   \authorrunning{Q. Wen, Y. Luo \& X. Kang }            %author_head in even pages
   \titlerunning{The formation of blue cluster in local Universe }  % title_head in odd pages

   \maketitle
%% The author head (on even pages) and the title head (on odd pages) will be
%% automatically extracted from \author{} and \title{}. Whenever the title is too long,
%% you will be asked to supply a shorter one by inserting either \authorrunning{} or
%% \titlerunning{} before \maketitle. Anyway, you can specify your own heads.
%%
%%
%% Note: In the following text body of your manuscript, please note several differences from
%%       other major journals:
%% (1) \subsection{Please Capitalize the First Letter of Each Notional Word in Subsection Title}
%% (2) Please Capitalize the First Letter of Each Notional Word in all tables' captions

%
%________________________________________________ sections below
%
\section{Introduction}           %% first-level sections will be auto-capitalized
\label{sect:intro}

In observation, blue galaxies have higher star formation rate and younger stellar population than red galaxies (\citealt{Kauffmann+etal+2003, Brinchmann+etal+2004}). There are a few physical mechanisms to transfer blue galaxies into red galaxies by reducing their star formation rate. For example, for central galaxies, AGN feedback can reduce the cooling rate of hot gas (\citealt{Henriques+etal+2019, Guo+etal+2011}) to stop the accumulation of cold gas. While for satellite galaxies, environment effect like gas stripping can lead to a rapid decline in star formation and reddening in color (\citealt{Baldry+etal+2006, Wang+etal+2007}). Therefore, a blue galaxy accreted into a cluster will become red gradually. This is known as the Butcher-Oemler effect (\citealt{Butcher+Oemler+1984}) that the number fraction of star-forming galaxies (blue fraction) in clusters reduces with decreasing redshifts and blue fraction of massive clusters is found to be less than 0.2 since $z<0.1$. 

Using the data from DEEP2, \citealt{Gerke+etal+2007} found that the blue fraction in groups at $0.75<z<1.0$ is around 0.6, which declines rapidly with decreasing redshifts. For example, \citealt{De Propris+etal+2004} found from the 2dFGRS data (\citealt{Colless+etal+2001}) that the blue fraction in groups is between 0.0 and 0.6 with a mean of 0.1 at $z\sim0$. Thus group or cluster with higher fraction of blue galaxies is not expected in the local Universe. Recently, \citealt{Hashimoto+etal+2019} (Hereafter H19) found a blue cluster with high fraction of blue galaxies ($0.57\pm0.06$) with halo mass around $2.0^{+1.9}_{-1.0}\times10^{14}M_{\odot}$ at $z\sim0.06$ in the SDSS DR7 (\citealt{Abazajian+etal+2009}). Actually, there has been no clear definition of  blue cluster in the literatures regarding to the fraction of blue galaxies in a cluster. In this paper, we follow the usual case to define a blue cluster with blue fraction larger than 0.5. Using filaments catalog from the SDSS (\citealt{Tempel+etal+2014}), H19 discovered that  this blue cluster is in a large-scale filamentary structure. They interpreted it as the evidence of environmental effect where the cold gas is accreted along filament to feed the blue galaxies. However, according to the simulation results (e.g., \citealt{Dekel+Birnboim+2006, Dekel+etal+2009}), the accretion of cold gas along filament will be shock heated and direct feeding is not expected in massive haloes ($>10^{12}M_{\odot}$). 

Although one particular blue cluster will not violate the general picture of galaxy evolution in cluster, it is interesting to investigate how this could happen and which mechanism is responsible for this particular blue cluster found by H19. Considering the extreme scarcity of blue cluster, it is also possible there is contamination in the algorithm of cluster finding. For example, H19 used the conventional friends-of-friends (FOF) algorithm to define a cluster. It is well known clusters found using the FOF algorithm depend strongly on the linking length and relative velocities between member galaxies (\citealt{Aguerri+etal+2007}). On the other hand, the theoretical predictions vary among the models. H19 compared the blue fraction with two semi-analytical models (SAMs), namely the GALACTICUS (\citealt{Benson+2012}) and $\nu^{2}$GC (\citealt{Ishiyama+etal+2015, Makiya+etal+2016}). It has been learned that the fraction of blue galaxies is under-predicted in most SAMs. Given the serendipity of this blue cluster and the uncertainty in modeling, one should take caution when comparing the data with theoretical predictions.

In this work, we use the group catalog constructed by \citealt{Yang+etal+2007} from the SDSS DR7 data to find blue clusters and compare the results with the prediction from the L-Galaxies (\citealt{Henriques+etal+2015}) which has been shown to well reproduce the fraction of red and blue galaxies in different haloes. Our main goal is to find what is the fraction of blue cluster in the Yang et al. catalog and how it is dependent on the group finding algorithm, and in particularly, we want to quantify if the formation of blue cluster is due to the effect of large-scale environment. The structure of this paper is as follows. In Section 2 we describe the selection of sample from the observation and the L-Galaxies model. Section 3 presents the statistical results from group catalog. In Section 4, we compare statistical properties of blue clusters with red clusters in the L-Galaxies. Then we conclude all results in Section 5. Throughout this paper, we adopt the Planck14 cosmology (\citealt{Planck+Collaboration+etal+2014}): ($\Omega_m, \Omega_{\Lambda}, \Omega_b, h$)=(0.315, 0.685, 0.0487, 0.673).  

%% Authors can give a citation as 'Michel et al. 1992'.
%% You may also use \cite, \citep and \citet for citation, and use Table~1 or Figure~1
%% and so forth. Using \ref and \label for cross-references of Tables/Figures
%% is a good way in adjusting/adding/removing text, tables or figures.

\section{Sample Selection}
\label{sect:Obs}

\subsection{SDSS DR7 group catalog}

We construct a volume-limited galaxy sample extracted from Sample C of SDSS DR7 group catalog (\citealt{Yang+etal+2007, Yang+etal+2008, Yang+etal+2009}) within $0.02<z<0.082$, which is same as H19. It is well known that some galaxies in SDSS do not have measured redshifts due to the fiber-collision effect. \citealt{Yang+etal+2007} have discussed this effect and argued that most of fiber-collision galaxies have redshifts within $500~km/s$ of their nearest neighbor (\citealt{Zehavi+etal+2002}). They assigned a fiber-collision galaxy the redshift of its nearest neighbor and also combined ROSAT X-ray cluster catalog. Hence, this effect may not affect our results significantly. In recent years, LAMOST survey has produced a complementary galaxy sample (\citealt{Luo+etal+2015, Shen+etal+2016, Zheng+Shen+2020}), and \citealt{Lim+etal+2017} modified the group finder of \citealt{Yang+etal+2007} with a $100\%$ redshift complementary galaxy sample. These efforts might reduce the impact of fiber-collision effect, but in this paper we still use the Yang et al. group catalog.

\begin{figure}
\centering
\includegraphics[width=8cm,angle=0]{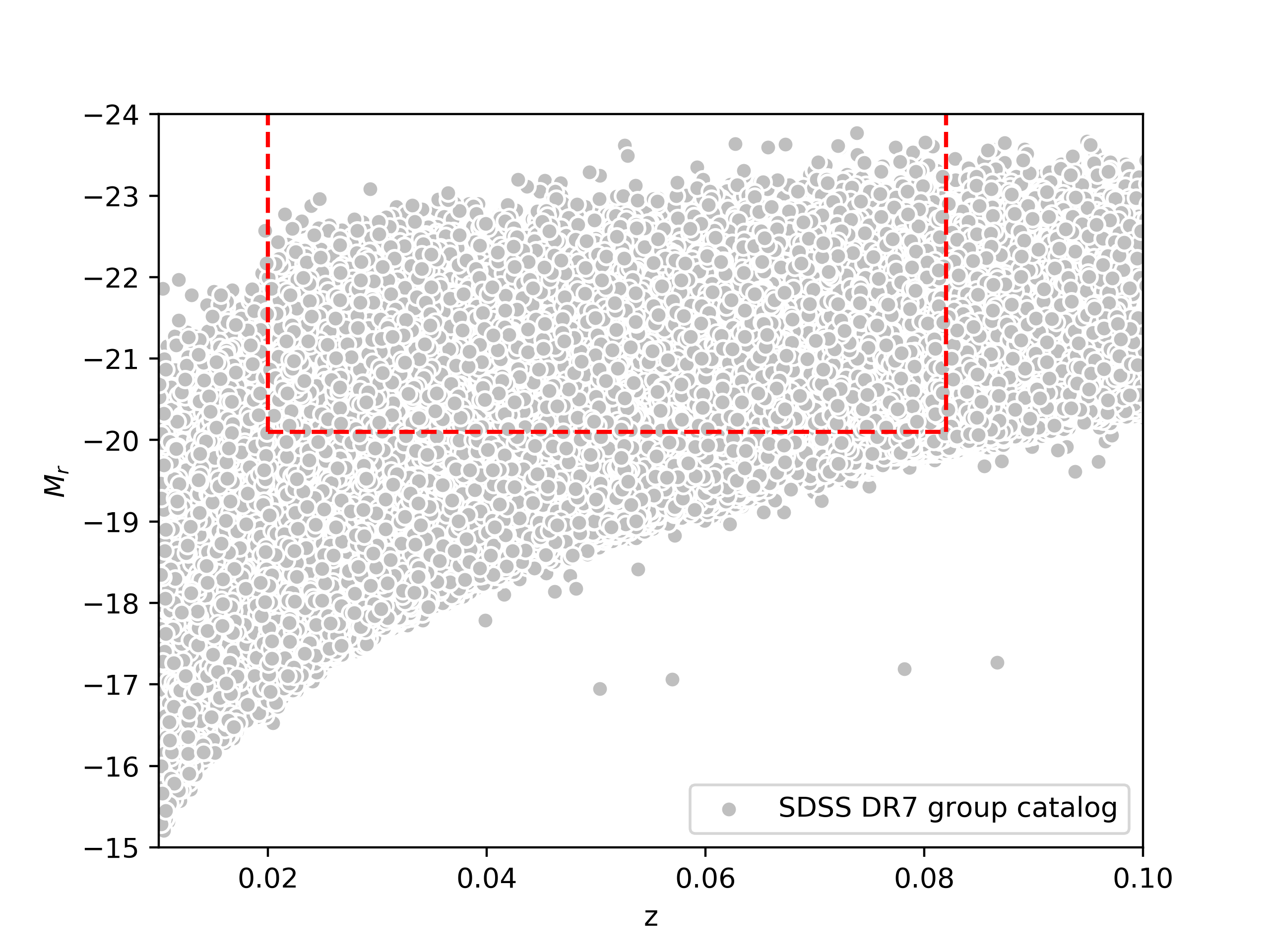}
\caption{\small The r-band absolute magnitude and redshift relation in the SDSS DR7 group catalog. The galaxies inside the red dashed region are selected as our volume-limited sample.}
\label{Fig 1}
\end{figure}

We get the group information and cluster halo mass from this group catalog. The stellar mass, magnitude of different bands and star formation rate are from MPA-JHU catalog. The morphology of galaxy is from catalog of \citealt{Tempel+etal+2011}. We exclude faint galaxies with r-band absolute magnitude, $M_r$, fainter than -20.1 mag, to ensure the completeness of the SDSS spectroscopic observations both for faint galaxies at higher z and brighter galaxies at lower z. Figure~\ref{Fig 1} shows the r-band absolute magnitude and redshift relation in the SDSS DR7 group catalog. The red dashed region represents the selected sample from group catalog for completeness. 

H19 apply the FOF algorithm to galaxies in high-density region with a projected linking length of 0.75 Mpc to find clusters, and they find member galaxies of cluster with a projected distance less than this linking length and velocity difference below 1000km/s. In fact a projected distance of 0.75 Mpc is roughly the projected virial radius of cluster with halo mass around $10^{14}M_{\odot}$. The choice of fixed linking length in H19 may have two effects. First, they may not be able to find cluster with mass larger than $10^{14}M_{\odot}$ or the massive cluster will be artificially split. Second, they may miss member galaxies at the outskirts of the cluster, which may be dominated by blue satellites. We will later discuss the limitation of their choice. Compered with FOF algorithm, the algorithm used by \citealt{Yang+etal+2007} to find group catalog performs much better (\citealt{Lim+etal+2017}). As a result, the number of members from our group catalog is usually larger than that from H19. 

\begin{figure}
\centering
\includegraphics[width=10cm,angle=0]{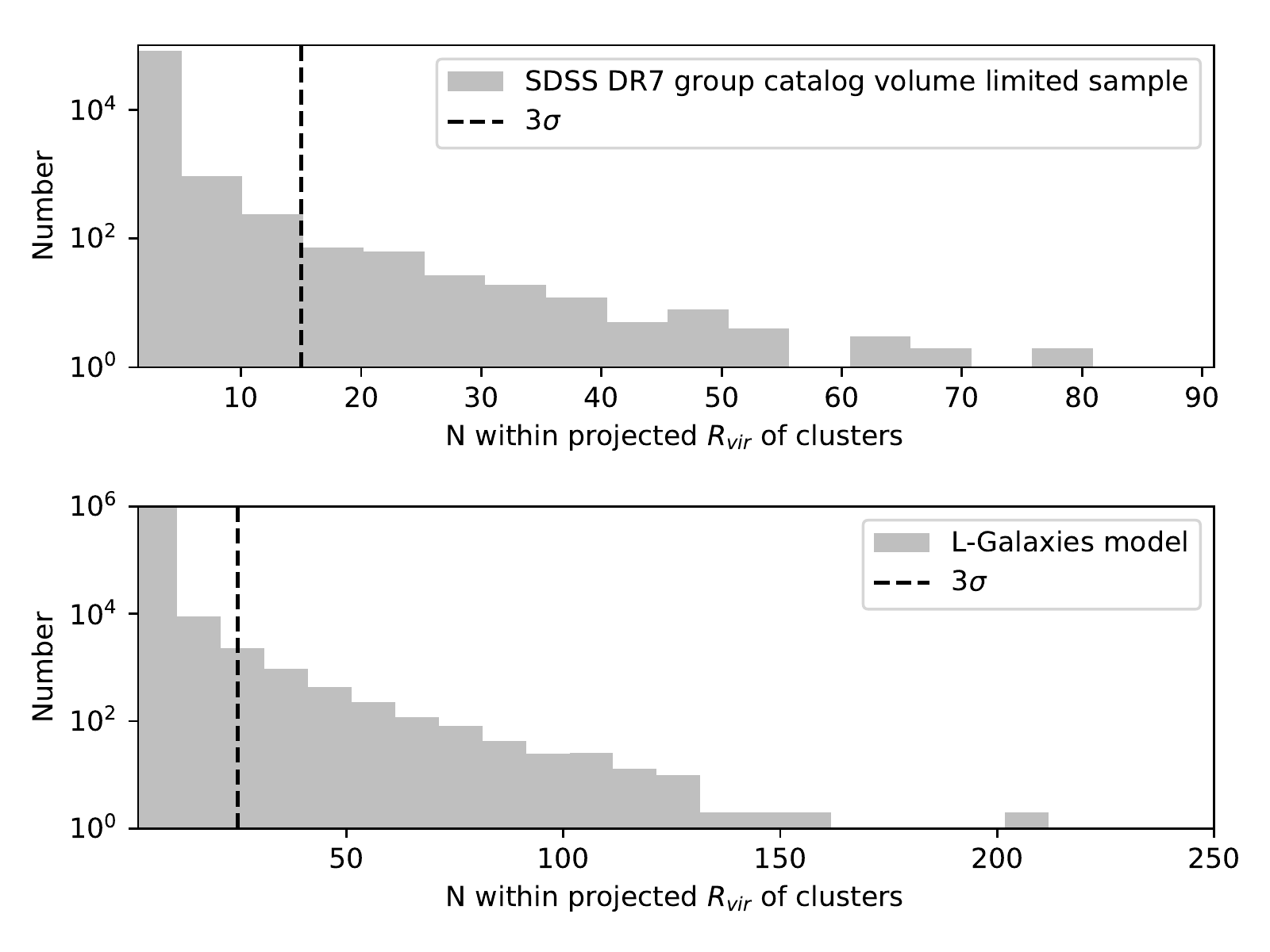}
\caption{\small A histogram of the number of galaxies within projected virial radius of clusters in SDSS DR7 group catalog (upper panel) and L-Galaxies (lower panel). The vertical dashed lines correspond to $3\sigma$ from the mean value of projected galaxy number of cluster. }
\label{Fig 2}
\end{figure}

In this paper, we calculate, $N$, the number of galaxies whose projected distance from the central galaxy are less than the projected virial radius of cluster, and with relative velocities to the central galaxy less than $1000~km/s$. As pointed by H19, $N$ is a good indicator of local environment of a cluster. In the upper panel of Figure~\ref{Fig 2}, we plot the histogram of $N$ from the SDSS DR7 group catalog. The vertical dashed line  is where the cluster environment density is above $3\sigma$ of the average environment density. In general, cluster in low-density region (lower than $3\sigma$) have member galaxies less than 15. 

To classify galaxies into red and blue, we use their specific star formation rate (sSFR). \citealt{Henriques+etal+2015} compared volume-limited sSFR distribution of L-Galaxies with that of SDSS galaxies in different stellar mass bins. The color of mock galaxies is slightly bluer than SDSS galaxies in some stellar mass bins. However, there is a clear bimodal distribution of sSFR in both L-Galaxies and SDSS, and the threshold value is about $10^{-11}~yr^{-1}$. In this paper we divide galaxies with sSFR larger than $10^{-11}~yr^{-1}$ as blue galaxies, to follow the popular threshold in literatures (\citealt{Mcgee+etal+2011, Hearin+etal+2015a}). In the following part of this paper, the blue galaxies are sometime called as star-forming galaxies and red galaxies as quenched galaxies. We also use $f_b$ to represent the number fraction of blue galaxies in each cluster.
% r-band Magnitude Mr-z relation

% number of galaxies within Rvir of clusters for SDSS DR7 group catalog and L-Galaxies

\subsection{L-Galaxies sample}

To construct the model group catalog, we use galaxy catalog at $z\sim0.05$ from the L-Galaxies sample which is implemented on Millennium Simulation (\citealt{Springel+etal+2005}). To be consistent with the analysis we have done for the SDSS group catalog, we implement a magnitude limit with $M_{r}$ = -20.1 mag and transfer the 3D physical distance to projected distance along the X axis, and use $V_x$ to calculate the relative velocity between two galaxies. Then we calculate the environment density of cluster in L-Galaxies as we have done in Yang group catalog. In lower panel of Figure~\ref{Fig 2} we plot the environment density distribution from the L-Galaxies model. As in the upper panel, the dashed line indicates the galaxy density is $3\sigma$ of the mean distribution. We note that at given environment density $N$, the model group catalog contains more clusters than the SDSS data. This is because our model galaxies are in a large-volume (box size $\sim$ 714 Mpc) than the volume-limited SDSS data. We do not normalize them to the same volume, as the distribution of SDSS volume-limited group catalog is more sparse in the space and being difficult to calculate the actual space volume. The large number of cluster from the model enable us to minimize the statistical uncertainty.

%in lower panel of  Figure~\ref{Fig 2} and select the clusters whose density is beyond $3\sigma$ deviated from the mean value of $N$.

\section{Results from SDSS DR7 group catalog}
\label{sect:SDSS}

To calculate the blue fraction of clusters and cross time more accurately, we exclude clusters whose number of member galaxies is less than 10. In the analysis of H19, the minimum halo mass of clusters in densest environment is about $10^{13.8}M_{\odot}$. We also remove clusters whose halo mass is less than $10^{13.8}M_{\odot}$. Here we also calculate the cross time $t_c$ for each cluster, defined as (\citealt{Ai+Zhu+2018, Tully+1987}), 
% cross time 
\begin{equation}
t_c=\frac{1.51^{1/2}R_{rms}}{3^{1/2}\sigma_P}
\label{eq:CrossTime}
\end{equation}
The parameter $t_cH_0$ represents the rough time that a galaxy crosses the cluster, and its reciprocal is the maximum number of times of a galaxy has crossed the cluster since forming (\citealt{Hickson+etal+1992}). $t_c$ can also be used to represent cluster age (\citealt{Diaferio+etal+1993}), and in general young clusters have large $t_c$ and vice versa. $R_{rms}$ is the rms projected radius of clusters (\citealt{Berlind+etal+2006}) and $\sigma_P$ is the projected velocity dispersion.

\begin{figure}
\centering
\includegraphics[width=10cm,angle=0]{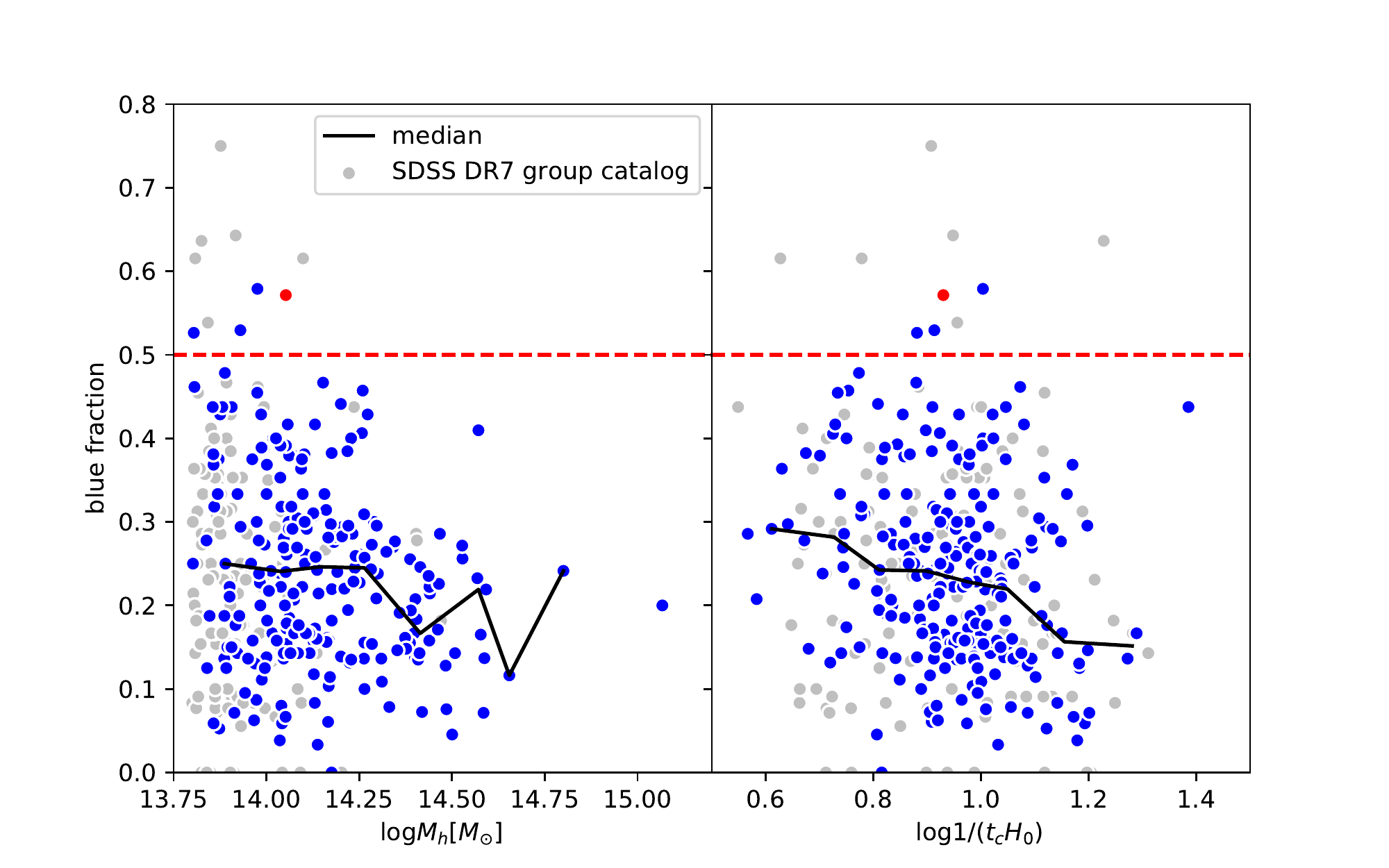}
\caption{\small The fraction of blue galaxies as a function of the cluster halo mass (left panel) and $1/(t_cH_0)$ (right panel) in SDSS DR7 group catalog. Gray and blue points are the clusters with number of member galaxies larger than 10 and 15 separately. Black solid line represents the median values, and red dashed line indicates that blue fraction of clusters is 0.5. The red point is the blue cluster found by H19.}
\label{Fig 3}
\end{figure}

Figure~\ref{Fig 3} shows the blue fraction of clusters as a function of halo mass and $1/(t_cH_0)$. The gray points are for all clusters with halo mass larger than $10^{13.8}M_{\odot}$ in all environment and blue points are for those only in high-density regions (with density higher than $3\sigma$ of the average). The horizontal red dashed line is the cluster with blue fraction of 0.5. Figure~\ref{Fig 3} shows that the blue fraction of clusters decreases with increasing halo mass and $1/(t_cH_0)$, consistent with other work (\citealt{Hashimoto+etal+2019, Wang+etal+2018, Baldry+etal+2006}). For clusters in high-density regions, we found 4 blue clusters whose blue fraction is larger than 0.5, one of them (red point) is the one found by H19. As the number of member galaxies of cluster in low-density region is lower than 15, the calculation of blue fraction has larger error, we will not discuss them in the following. Compared with H19, we found more blue clusters in high-density region. We note that H19 did not find the other three blue clusters in our work, it is not because they applied a mass cut to their catalog. They applied their cluster finding algorithm to galaxies in high-density region which leads to halo mass with a lower limit of $10^{13.8}M_{\odot}$. We also apply the cluster finding algorithm of H19 to the four blue clusters in our sample, and find that they all have mass larger than $10^{13.8}M_{\odot}$, similar to the halo mass in Yang et al. catalog. Thus H19 failed to find more blue clusters mainly due to their fixed linking length which is too small for massive clusters, not the halo mass cut in their analysis.

% blue fraction in SDSS DR7 group catalog

\begin{figure}
\centering
\includegraphics[width=15cm,angle=0]{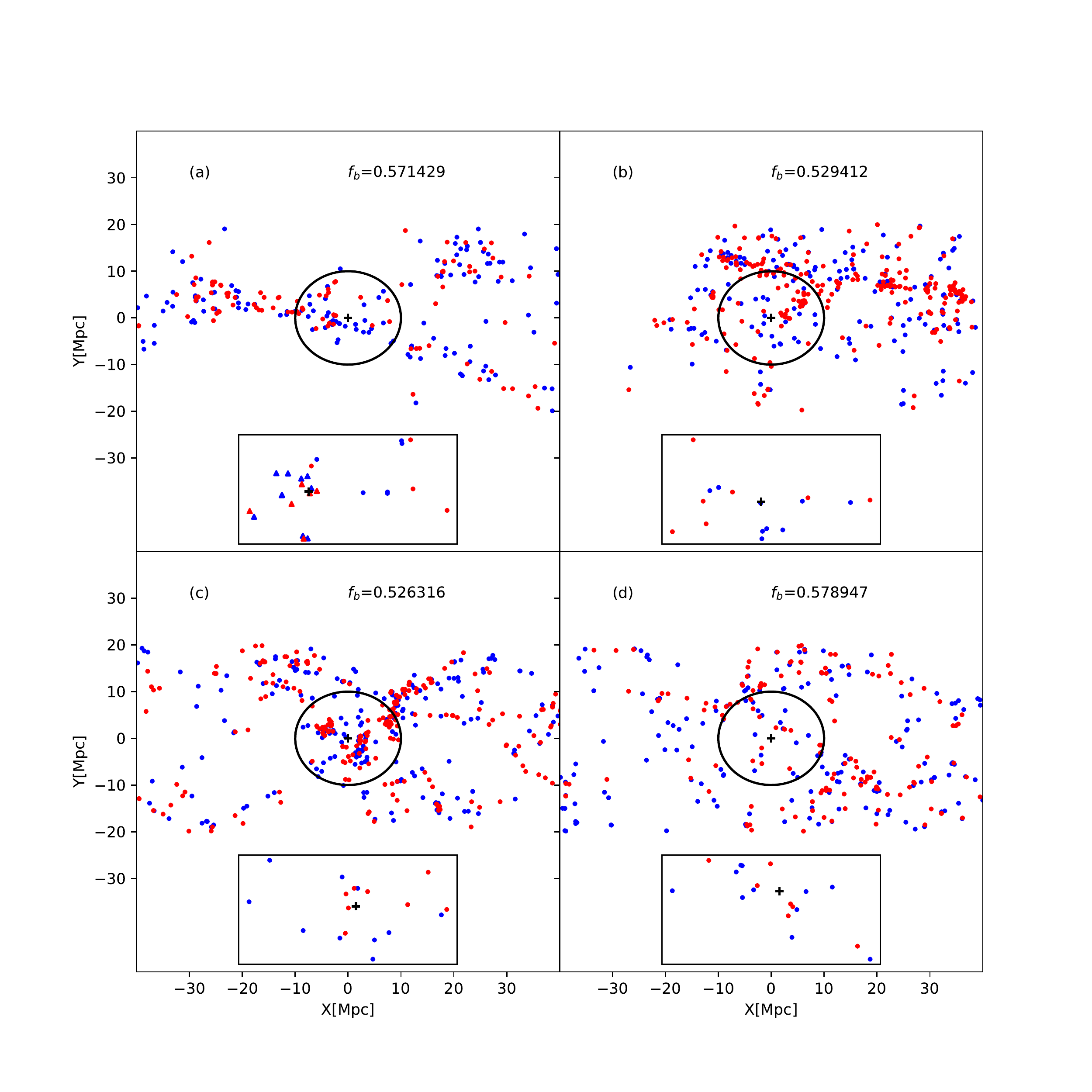}
\caption{\small Large scale structure map for four clusters whose fraction of blue member galaxies is larger than 0.5. The cluster in panel (a) is found by H19. Red and blue points correspond to blue and red galaxies respectively. The region of 10 Mpc around of cluster center (black cross) is indicated by a black circle. We note the cluster member galaxies are not shown in this region. The inner box (represents $2.5\times1.5$ Mpc region) of each panel shows the member galaxies distribution of cluster, the triangles indicate galaxies found by H19.}
\label{Fig 4}
\end{figure}

We select these four blue clusters to examine their large-scale environment in Figure~\ref{Fig 4}. The upper left panel is for the blue cluster found by H19. Here we plot the spatial distribution of galaxies within 40 Mpc from these blue clusters (but member galaxies of the clusters are not shown). In each panel, we use black cross to represent the cluster center that is the median values of coordinates of member galaxies. The red and blue points are for red and blue galaxies respectively. The black circles denote the region with 10 Mpc of radius around the cluster center and all member galaxies of the cluster are shown in the inserted panel. The large-scale galaxy distribution around blue cluster in Panel (a) shows a filamentary structure and the cluster is located at the intersection of two filaments. The fraction of blue galaxies (except cluster members) within 10 Mpc around this cluster is larger than 0.6, while the other three clusters do not show such high fraction (about 0.3$\sim$0.5). This large-scale galaxies distribution is similar to that found by H19, and seems to support their argument that this blue cluster is feeding by more infalling blue galaxies or gas accretion along the filament. However, by checking the other three blue clusters, we do not find any filamentary structure around, neither any preferential blue galaxies on large scales. Thus, we conclude that not all blue clusters are formed due to the large-scale environment. 

% large scale structure map in SDSS DR7 group catalog

\begin{figure}
\centering
\includegraphics[width=10cm,angle=0]{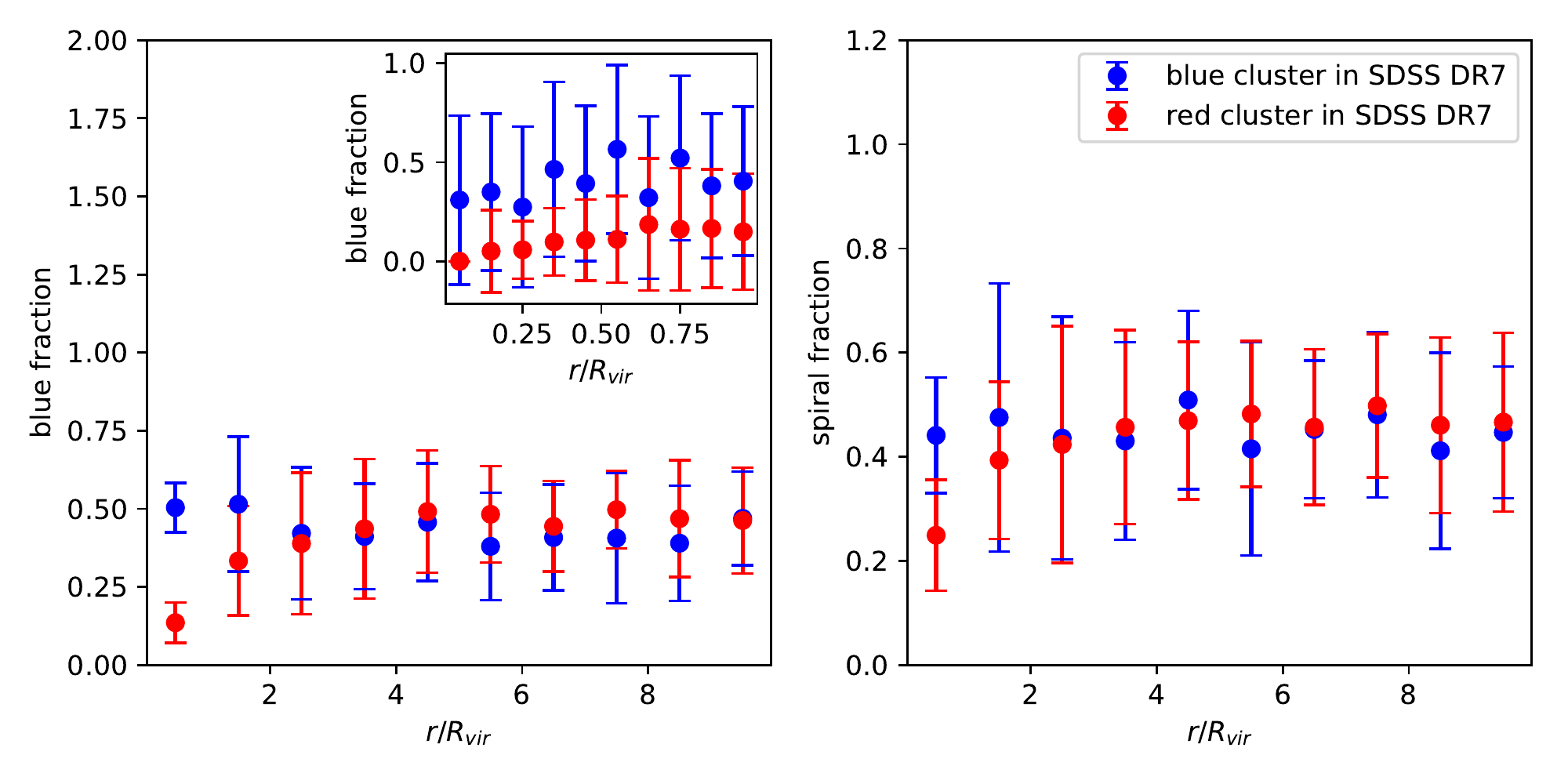}
\caption{\small Mean fraction of blue galaxies (left panel) and spiral galaxies (right panel) for two kinds of clusters as a function of specific projected radius in SDSS DR7 group catalog. The morphology information of galaxies is from catalog of \citealt{Tempel+etal+2011}. The blue and red points represent mean value of blue and red clusters respectively. The error bar is $1\sigma$ deviated from mean values. The inner figure of left panel is the zoomed results within projected virial radius of clusters.}
\label{Fig 5}
\end{figure}

The four blue clusters have similar halo mass between $13.8 <\log{(M_h/M_{\odot})}<14.2$. To investigate their properties in detail, we select all clusters within this mass range and divide them into blue clusters with $f_b\ge0.4$ and red clusters with $f_b\le0.2$. Note that here we use $f_b=0.4$ rather $0.5$ to divide clusters into blue clusters, purely for the purpose of increasing the sample size. We have tested that our results are not significantly affected by this choice. We plot the mean fraction of blue galaxies for the two samples as a function of projected distance from the cluster center in the left panel of Figure~\ref{Fig 5}, and we show the mean fraction of spiral galaxies in the right panel. We find that blue fraction in red clusters is less than blue clusters within $2R_{vir}$, while they have similar values on large scale. This result is consistent with other work (\citealt{Bray+etal+2016, Hearin+etal+2015a}), which found that the quench of galaxies basically happens on scales around the size of the cluster virial radius. The same trend is also seen for the fraction of spiral galaxies in blue and red cluster, though the effect is more mild.

We zoom in the radial distribution of blue fraction in the inserted panel in left panel of Figure~\ref{Fig 5}. It is seen that the fraction of blue galaxies increases with the distance to the cluster center,  showing that blue galaxies reside preferentially in outskirts of  the cluster. \citealt{Wetzel+etal+2012} also found similar result that the fraction of quenched galaxies decreases as projected radius within virial radius of clusters in SDSS DR7 group catalog. As mentioned in Section 2, FOF method may not find member galaxies that live in outskirt of cluster, thus it has high probability to miss the blue galaxies at cluster outskirts. In particular, we show the member galaxies distribution of four blue clusters in Figure~\ref{Fig 4}, the triangles of panel (a) represent member galaxies found by H19, we note that missed galaxies reside in outskirt of this cluster. This also explains why H19 just found one blue cluster whose blue fraction is larger than 0.5 while we find 4 from the our group catalog.

\begin{figure}
\centering
\includegraphics[width=15cm,angle=0]{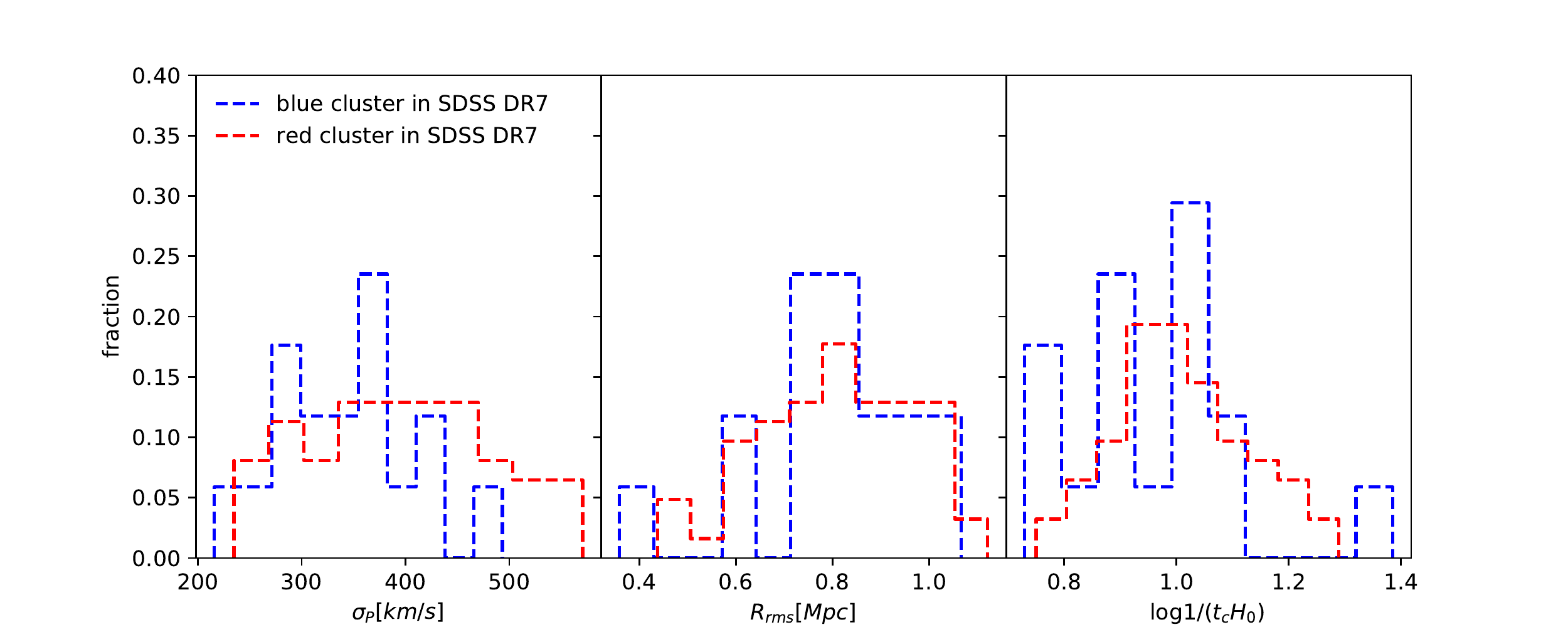}
\caption{\small Projected velocity dispersion $\sigma_P$ (left panel), rms projected radius $R_{rms}$ (middle panel) and $1/(t_cH_0)$ distribution (right panel) for blue clusters (blue dashed line) and red clusters (red dashed line) in SDSS DR7 group catalog.}
\label{Fig 6}
\end{figure}

We also compare the distributions of projected velocity dispersion $\sigma_P$, the rms projected radius $R_{rms}$ and $1/(t_cH_0)$ for blue and red clusters in Figure~\ref{Fig 6}. It is found these clusters with similar halo mass have similar distribution of $R_{rms}$, but blue clusters have lower $\sigma_P$, thus lower $1/(t_cH_0)$ (see Equation.1). As mentioned above, $1/(t_cH_0)$ can be as an indicator of cluster age. This result suggests that blue clusters formed later than red clusters. These results are also consistent with previous work (\citealt{Hashimoto+etal+2019, Aguerri+etal+2007}).  

% statistical filaments fraction and spirals fraction of clusters in SDSS DR7 group catalog

% spiral fraction and cross time distribution in SDSS DR7 group catalog

\section{Results from L-Galaxies model}
\label{sect:lgalaxies}
\begin{figure}
\centering
\includegraphics[width=10cm,angle=0]{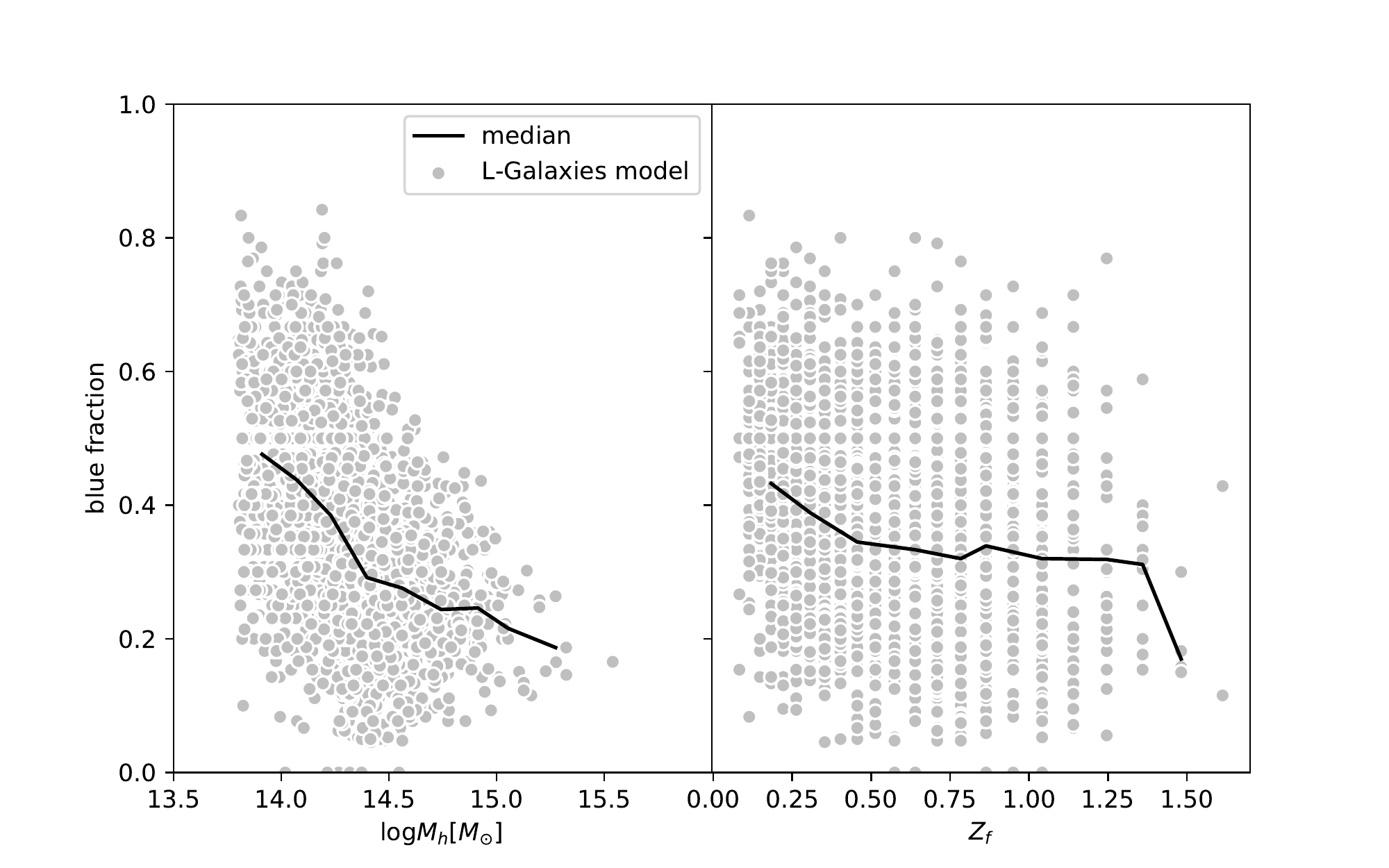}
\caption{\small The fraction of blue galaxies as a function of halo mass (left panel) and formation time (right panel) in L-Galaxies. Black solid line represents median values.}
\label{Fig 7}
\end{figure}

Now we show the results from the mock catalog constructed using the L-Galaxies model. As we have done for observational data, we exclude clusters whose number of member galaxies less than 10 and halo mass lower than $10^{13.8}M_{\odot}$. We plot the blue fraction of clusters as a function of halo mass and formation time in Figure~\ref{Fig 7}. Usually the formation time, $Z_f$, is the time when the main branch of halo merger tree has assembled half of its present mass (\citealt{Tojeiro+etal+2017, Li+etal+2008}), which is a good indicator of cluster age (\citealt{Shi+etal+2018, Tojeiro+etal+2017}) that old clusters have large formation times and vice versa. We find that blue fraction declines with halo mass and formation time increasing, which is consistent with observation shown in Figure~\ref{Fig 3}. We note that the L-Galaxies model seems to predict higher fraction of blue cluster than the observational data. This is because the model slightly over-predicts the fraction of blue galaxies in low-mass galaxies, as shown in \citealt{Henriques+etal+2015}.

\begin{figure}
\centering
\includegraphics[width=8cm,angle=0]{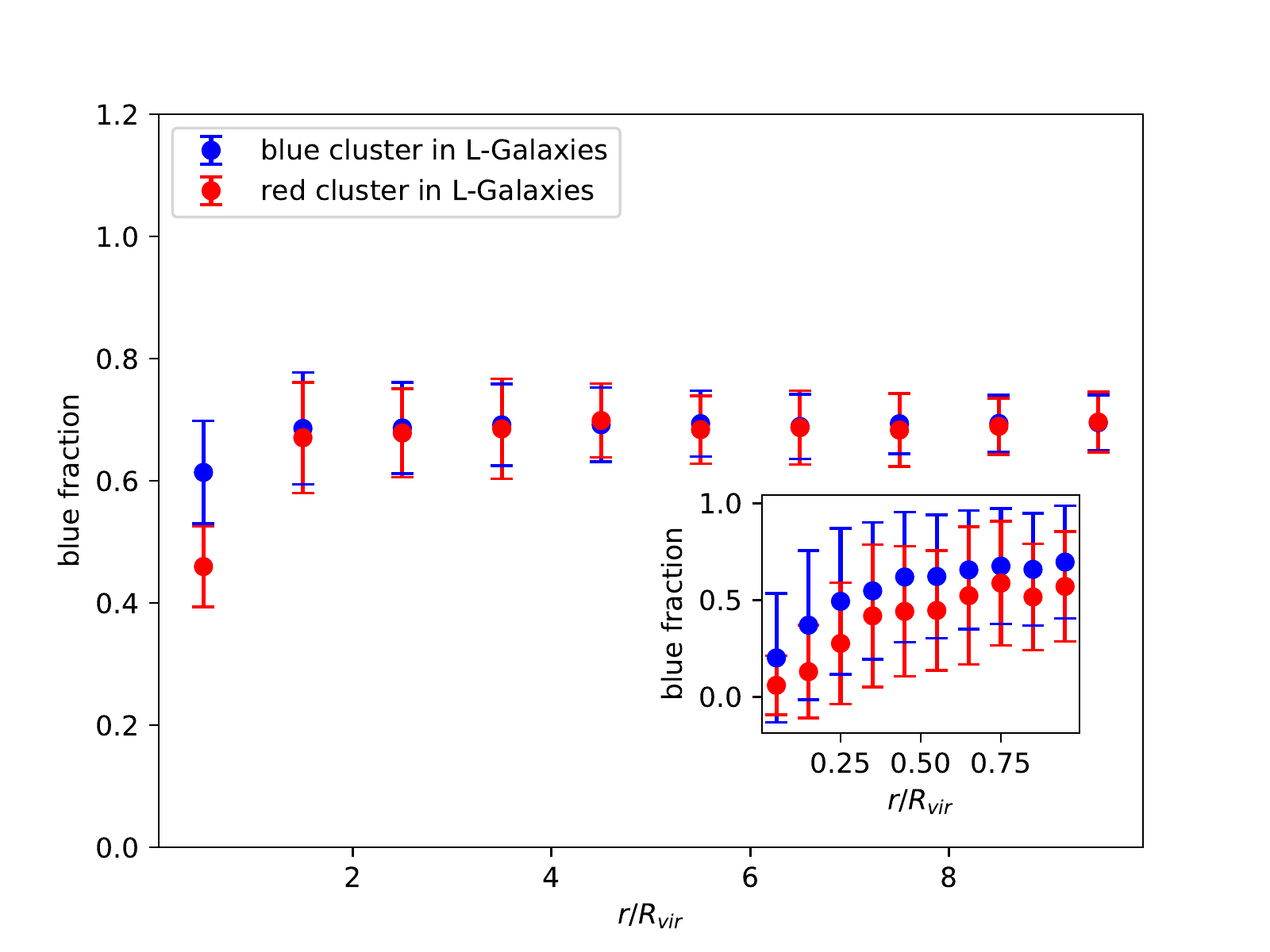}
\caption{\small Mean fraction of blue galaxies as a function of specific projected radius in L-Galaxies. The blue and red points represent mean value of blue and red clusters respectively. The error bar is $1\sigma$ deviated from mean values. The inner figure is the zoomed results within projected $R_{vir}$ of clusters.}
\label{Fig 8}
\end{figure}

We also show the mean fraction of blue galaxies in blue and red clusters as a function of specific projected radius in Figure~\ref{Fig 8}. Again here the clusters whose halo mass is between $10^{13.8}M_{\odot}$ and $10^{14.2}M_{\odot}$ are divided into blue clusters with $f_b\ge0.4$, and $f_b\le0.2$ with red clusters. We find the similar results as observation (see left panel of Figure~\ref{Fig 5}), which confirms that star-forming galaxies reside preferentially in clusters outskirts and blue fraction of clusters is affected mainly by local environments. 

\begin{figure}
\centering
\includegraphics[width=15cm,angle=0]{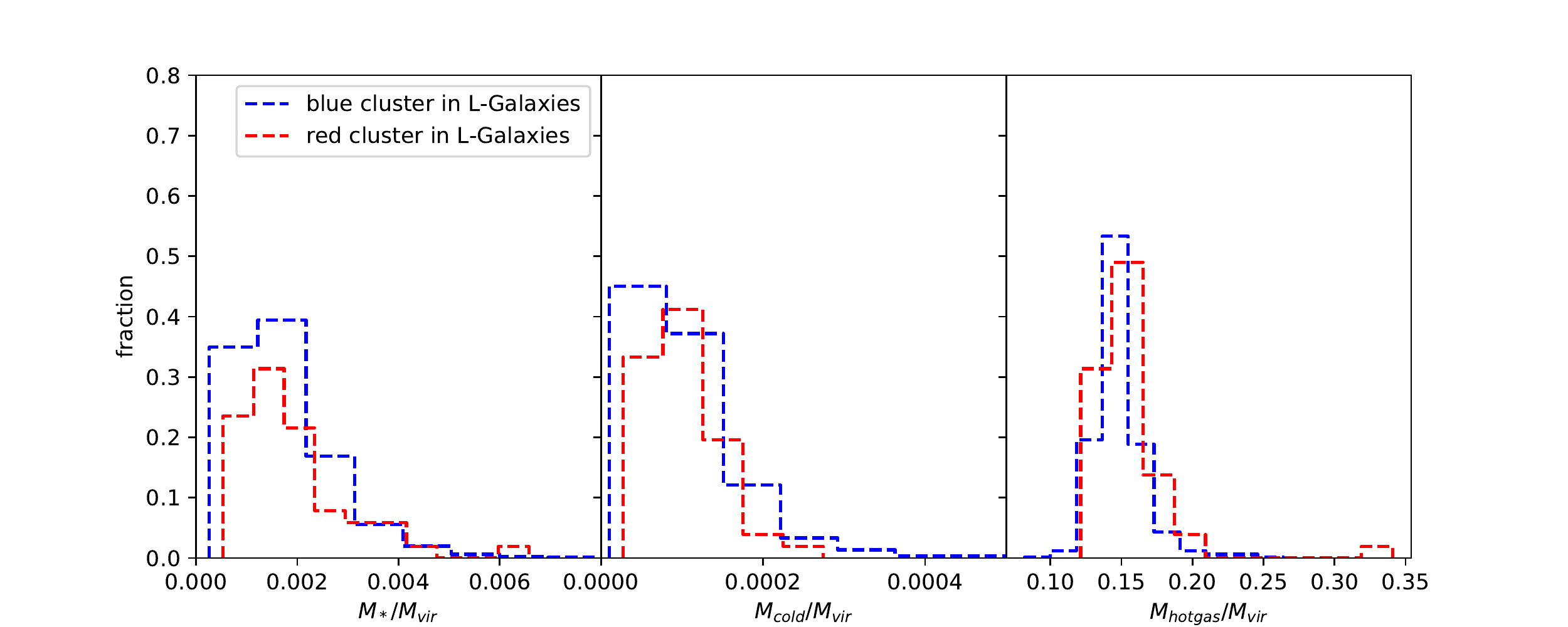}
\caption{\small Stellar mass fraction (left panel), cold gas fraction (middle panel) and hot gas fraction (right panel) distribution of central galaxies for two kinds of clusters in L-Galaxies. The red and blue dashed lines represent red and blue clusters respectively.}
\label{Fig 9}
\end{figure}

\begin{figure}
\centering
\includegraphics[width=15cm,angle=0]{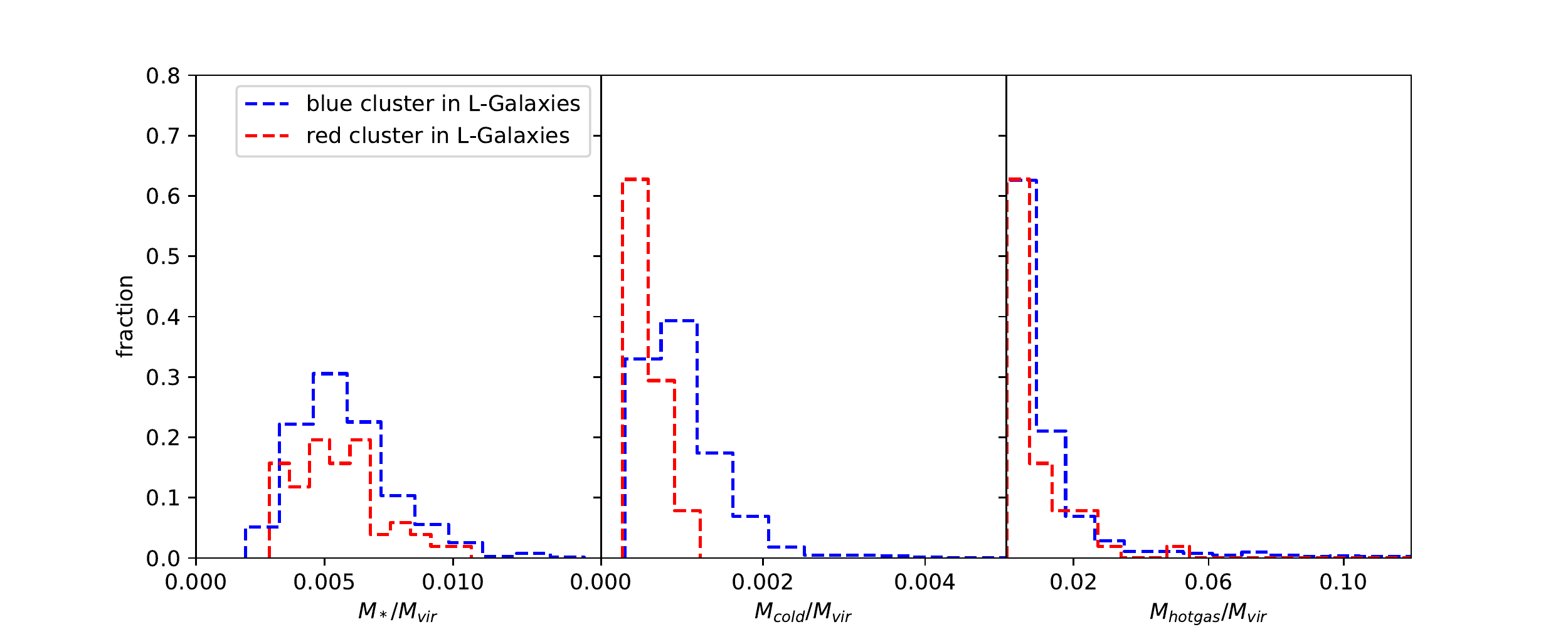}
\caption{\small Same as Figure~\ref{Fig 9} but for satellite galaxies for two kinds of clusters.}
\label{Fig 10}
\end{figure}

\begin{figure}
\centering
\includegraphics[width=10cm,angle=0]{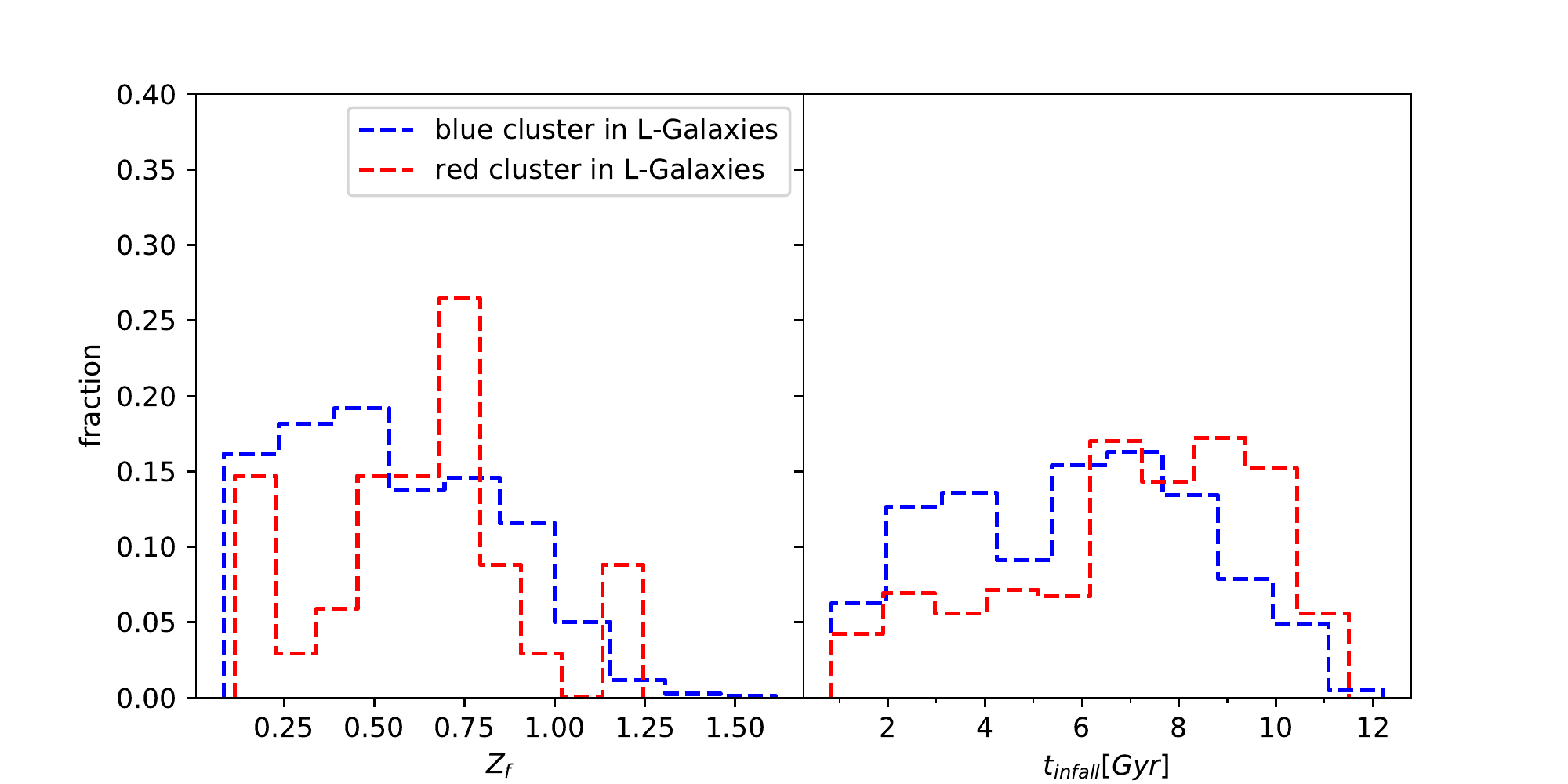}
\caption{\small Formation time of clusters distribution (left panel) and infalling time of satellite galaxies distribution (right panel) in L-Galaxies. The blue and red dashed lines represent blue and clusters respectively.}
\label{Fig 11}
\end{figure}

Up to now, we did not distinguish the satellite galaxies from the centrals. To see which type of galaxies contributes most to the blue fraction, we show their stellar mass distribution, hot gas distribution and cold gas mass distribution in Figure~\ref{Fig 9} and Figure~\ref{Fig 10}, for central and satellite galaxy respectively. We find that baryon fraction of central galaxies is similar in both blue and red clusters, while the cold gas of satellite galaxies in blue clusters is higher than those in red clusters. This indicates that the blue clusters is mainly contributed by star-forming and gas-rich satellite galaxies. In Figure~\ref{Fig 11}, we plot the formation time distribution and the accretion time distribution for satellites in the clusters. It is seen that the formation time of blue clusters is lower than red clusters and the infalling time of satellites in blue clusters is lower, which suggests that there are more recently accreted  satellite galaxies in blue clusters, and most satellites retain enough cold gas to feed their star formation and being blue.

% blue fraction in L-Galaxies galaxy catalog

% statistical mean quench fraction of clusters in L-Galaxies

% statistical baryon fraction of clusters for central galaxy in L-Galaxies

% statistical baryon fraction of clusters for satellite galaxy in L-Galaxies

% statistical formation time of clusters in L-Galaxies

\section{Conclusions}
\label{sect:conclusion}

We construct a volume-limited sample from SDSS DR7 group catalog at $0.02<z<0.082$ and study the fraction of blue galaxies in clusters in high-density regions. We also compare the data with the L-Galaxies model to investigate the origin of the blue clusters. Our main findings are:

\begin{enumerate}
\item The fraction of blue galaxies decreases with increasing halo mass and cluster age in both observation and model, which are both consistent with the H19 results.
\item The high fraction of blue galaxies in blue clusters is mainly from galaxies at distance lower than projected $2R_{vir}$. On large scales, red cluster and blue cluster have similar fraction of blue galaxies. Such an effect is also seen in both data and model.
\item We find four blue clusters from the SDSS DR7 group catalog in local Universe, with the fraction of blue galaxies larger than 0.5. One cluster as found by H19 is in a filamentary structure, the other three blue clusters show no clear filamentary structure around.
\item Using the L-Galaxies model, we find that blue clusters form later than red clusters, with more recently accreted blue satellite galaxies at the outskirts of the clusters.
\end{enumerate}

We conclude that the blue cluster found by H19 is indeed in a filamentary structure, indicating possible large-scale environmental effect. However, the three additional blue clusters found in our work do not live in filamentary structures and our work suggests that the formation of blue clusters is mainly a local effect as a result of more newly accreted star forming satellite galaxies. 

\begin{acknowledgements}
We thank Lei Wang for providing the SDSS DR7 group catalog and thank the anonymous referee for useful comments and suggestions to improve the manuscript. This work used the 2015 public version of the Munich model of galaxy formation and evolution: L-Galaxies. The source code and a full description of the model are available at https://lgalaxiespublicrelease.github.io/. The Millennium Simulation data bases used in this article and the web application providing online access to them were constructed as part of the activities of the German Astrophysical Virtual Observatory (GAVO). This work was funded by the 973 program (No. 2015CB857003), the NSFC (No. 11825303, 11861131006, 11333008, 11703091).
\end{acknowledgements}

\label{lastpage}


\begin{thebibliography}{99}
%% you can type \apj for ApJ, \aap for A&A, \apss for Ap&SS, etc. Please consult
%% the macro chjaa.cls. You can also find them in aasguide.tex (AASTeX for ApJ, AJ, PASP)
%% Please follow the format of ChJAA's reference list
	
	\bibitem[Abazajian et al.(2009)]{Abazajian+etal+2009} Abazajian K. N. et al., 2009, ApJS, 182, 543 
	
	\bibitem[Aguerri et al.(2007)]{Aguerri+etal+2007} Aguerri J. A. L. et al., 2007, \aap, 471, 17
	
	\bibitem[Ai \& Zhu (2018)]{Ai+Zhu+2018} Ai M., Zhu M., 2018, ApJ, 862, 48
	
	\bibitem[Baldry et al.(2006)]{Baldry+etal+2006} Baldry I. K., Balogh M. L., Bower R. G. et al., 2006, MNRAS, 373, 469
	
	\bibitem[Benson (2012)]{Benson+2012} Benson A. J., 2012, New Astron., 17, 175
	
	\bibitem[Berlind et al.(2006)]{Berlind+etal+2006} Berlind A. A. et al., 2006, ApJS, 167, 1
	
	\bibitem[Bray et al.(2016)]{Bray+etal+2016} Bray A.D., Pillepich A. et al., 2016, MNRAS, 455, 185
	
	\bibitem[Brinchmann et al.(2004)]{Brinchmann+etal+2004} Brinchmann J., Charlot S. et al., 2004, MNRAS, 351, 1151
	
	\bibitem[Butcher \& Oemler (1984)]{Butcher+Oemler+1984} Butcher H., Oemler Jr. A., 1984, ApJ, 285, 426
	
	\bibitem[Colless et al.(2001)]{Colless+etal+2001} Colless M. et al., 2001, MNRAS, 328, 1039 
	
	\bibitem[Dekel \& Birnboim (2006)]{Dekel+Birnboim+2006} Dekel A., Birnboim Y., 2006, MNRAS, 368, 2
	
	\bibitem[Dekel et al.(2009)]{Dekel+etal+2009} Dekel A., Birnboim Y. et al., 2009, Nature, 457, 451
	
	\bibitem[De Propris et al.(2004)]{De Propris+etal+2004} De Propris R. et al., 2004, MNRAS, 351, 125
	
	\bibitem[Diaferio et al.(1993)]{Diaferio+etal+1993} Diaferio A., Ramela M. et al., 1993, AJ, 105, 2035
	
	\bibitem[Gerke et al.(2007)]{Gerke+etal+2007} Gerke B. F., Newman J. A. et al., 2007, MNRAS, 376, 1425
	
	\bibitem[Guo et al.(2011)]{Guo+etal+2011} Guo Q., White S. D. M., Boylan-Kolchin M. et al., 2011, EAS, 48, 447
	
	\bibitem[Hashimoto et al.(2019)]{Hashimoto+etal+2019} Hashimoto T., Goto T. et al., 2019, MNRAS, 489, 2014
	
	\bibitem[Hearin et al.(2015a)]{Hearin+etal+2015a} Hearin A.P., Watson D.F. et al., 2015a, MNRAS, 452, 1958

	\bibitem[Henriques et al.(2015)]{Henriques+etal+2015} Henriques B. M. B., White S. D. M., Thomas P. A. et al., 2015, MNRAS, 451, 2663
	
	\bibitem[Henriques et al.(2019)]{Henriques+etal+2019} Henriques B. M. B., White S. D. M., Lilly S. J. et al., 2019, MNRAS, 485, 3446
	
	\bibitem[Hickson et al.(1992)]{Hickson+etal+1992} Hickson P., Oliveira D. et al., 1992, ApJ, 399, 353H
	
	\bibitem[Ishiyama et al.(2015)]{Ishiyama+etal+2015} Ishiyama T., Enoki M. et al., 2015, PASJ, 67, 61
	
	\bibitem[Kauffmann et al.(2003)]{Kauffmann+etal+2003} Kauffmann G. et al., 2003, MNRAS, 341, 54
	
	\bibitem[Li et al.(2008)]{Li+etal+2008} Li Y., Mo H. J., Gao L., 2008, MNRAS, 389, 1419
	
	\bibitem[Lim et al.(2017)]{Lim+etal+2017} Lim S. H., Mo H. J. et al., 2017, MNRAS, 470, 2982 
	
	\bibitem[Luo et al.(2015)]{Luo+etal+2015} Luo A.-L. et al., 2015, RAA, 15, 1095
	
	\bibitem[Makiya et al.(2016)]{Makiya+etal+2016} Makiya R. et al., 2016, PASJ, 68, 25
	
	\bibitem[Mcgee et al.(2011)]{Mcgee+etal+2011} Mcgee S. L., Balogh M. L. et al., 2011, MNRAS, 413, 996
	
	\bibitem[Planck Collaboration et al.(2014)]{Planck+Collaboration+etal+2014} Planck Collaboration XVI, 2014, \aap, 571, A16
	
	\bibitem[Shen et al.(2016)]{Shen+etal+2016} Shen S.-Y. et al., 2016, RAA, 16, 43
	
	\bibitem[Shi et al.(2018)]{Shi+etal+2018} Shi J., Wang H. et al., 2018, ApJ, 857, 127
	
	\bibitem[Springel et al.(2005)]{Springel+etal+2005} Springel V., White S. D. M. et al., 2005, Nat., 435, 629
	
	\bibitem[Tempel et al.(2011)]{Tempel+etal+2011} Tempel E., Saar E. et al., 2011, \aap, 529, A53
	
	\bibitem[Tempel et al.(2014)]{Tempel+etal+2014} Tempel E., Stoica R. S. et al., 2014, MNRAS, 438, 3456
	
	\bibitem[Tojeiro et al.(2017)]{Tojeiro+etal+2017} Tojeiro R., Eardley E. et al., 2017, MNRAS, 470, 3720
	
	\bibitem[Tully (1987)]{Tully+1987} Tully R.B., 1987, ApJ, 321, 280
	
	\bibitem[Wang et al.(2007)]{Wang+etal+2007} Wang L., Li C. et al., 2007, MNRAS, 377, 1419
	
	\bibitem[Wang et al.(2018)]{Wang+etal+2018} Wang H. et al., 2018, ApJ, 852, 31
	
	\bibitem[Wetzel et al.(2012)]{Wetzel+etal+2012} Wetzel Andrew R. et al., 2012, MNRAS, 424, 232
	
	\bibitem[Yang et al.(2007)]{Yang+etal+2007} Yang X., Mo H. J. et al., 2007, ApJ, 671, 153

	\bibitem[Yang et al.(2008)]{Yang+etal+2008} Yang X., Mo H. J. et al., 2008, ApJ, 676, 248
	
	\bibitem[Yang et al.(2009)]{Yang+etal+2009} Yang X., Mo H. J. et al., 2009, ApJ, 695, 900
	
	\bibitem[Zehavi et al.(2002)]{Zehavi+etal+2002} Zehavi I. et al., 2002, ApJ, 571, 172
	
	\bibitem[Zheng \& Shen (2020)]{Zheng+Shen+2020} Zheng Y.-L., Shen S.-Y., 2020, ApJS, 246, 12Z
	
	

\end{thebibliography}
\end{document}